# PHASE SHIFT ANALYSIS OF ELASTIC $n^3He$ SCATTERING

S. B. Dubovichenko[1], Ch.T. Omarov[1]

[1] *Fessenkov Astrophysical Institute "NCSRT" NSA RK, 050020, Almaty, Kazakhstan,*
*dubovichenko@mail.ru, http://dubovichenko.ru*

N.A. Burkova [2]

[2] *Al-Farabi Kazakh National University 050000, Almaty, Kazakhstan*
*natali.burkova@gmail.com*

Basing on the kwon experimental measurements of differential cross sections on $n^3$He elastic scattering in the angular energy range $20^0$ - $160^0$ the standard phase shifts analysis was performed at the energies from 1 MeV up to 5 MeV.

*Key words*: light atomic nuclei, low and astrophysical energies, elastic scattering, $n^3$He system, phase shifts analysis

## Introduction

In [1] we reported on the possibility of description of Coulomb form-factors of lithium isotopes within the framework of potential cluster model (PCM) [2,3] based on the deep attractive potentials including forbidden by Pauli principle states [4] revealed in two- body cluster interaction potentials [5]. So then we succeeded in the reproducing of all available data on $^6$Li [6] characteristics including very queasy one, i.e. quadrupole momentum calculated in PCM with tensor forces. Following this approach we succeeded in the description of the astrophysical *S*-factor of the radiative capture in the following systems as $^2$H$^3$He, $^3$He$^4$He, $^3$H$^4$He, $^2$H$^4$He, $n^7$Li, $n^2$H, $p^2$H, and $p^{12}$C [7], but basing on the modified PCM (MPCM), see for details [8].

Actual results for MPCM we expert as the essential ansatz of the declared method [3,9] of the constructing of the interaction potentials according the symmetry classification by Young schemes. One of the critical characteristics might be recognized the astrophysical *S* -factor [10], or the coordinating total cross sections of the corresponding reactions [11], as well as radiative $n^3$He capture at astrophysical energies treated in our earlier works [3,4,7,8].

Thus, we are going to apply this very method to the description of the $n^3$He scattering channel bearing in mind quite analogue similar phase shift analysis for the following systems $p^6$Li [12], $n^{12}$C [13], $p^{12}$C [14], $^4$He$^4$He [15], $^4$He$^{12}$C [16], $p^{13}$C [17], $p^{14}$C [18] и $n^{16}$O [19] rather following the purposes of astrophysical applications.

## Phase shifts analysis methods

Just following the compact presentation of the given calculations let us remark that elastic scattering cross sections in the systems with spin structure 1/2 + 1/2 have more or less simple form, but cumbersome to somewhat extent, so reasonable refer to [20, 21]. Data on the experimental differential cross sections for the elastic scattering allows to reproduce the corresponding set of phase shifts $\delta_L^J$ with certain accuracy.

The quality of the theoretical model and experimental data coordination are presented according the well known relation

$$\chi^2 = \frac{1}{N}\sum_{i=1}^{N}\left[\frac{\sigma_i^t(\theta)-\sigma_i^e(\theta)}{\Delta\sigma_i^e(\theta)}\right]^2 = \frac{1}{N}\sum_{i=1}^{N}\chi_i^2 ,$$

where $\sigma^e$ and $\sigma^t$ are the experimental cross section and theoretical one, i.e. calculated cross section at some given values of the elastic phase shifts $\delta_L^J$ for the definite *i* number of the scattering angles, $\Delta\sigma^e$ – is the error of experimental cross sections measured at fixed angles, and *N* is a number of experimental dots.

The details for the computing program for the calculations of the total and differential cross sections with the half-integer spin, as well as for the phase shift analysis are given in [21,22]. To computing the corresponding program has been converted into Fortran 90 with double accuracy [3]. The speed calculation has been raised up essentially. Thus, we reached the accuracy $\varepsilon = (\chi_i^2 - \chi_{i-1}^2)$ within the $10^{-10}$ at the *i*-th step of the minimization procedure, and the calculation time was reduced factor of order comparing the analogue programs done in Turbo Basic, Borland Inc. [21,22].

We should remark that all masses have been taken as integer ($m_n = 1$, $m_{He} = 3$), and the recalculating constant was $\hbar / m_0 = 41.4686$ MeV fm$^2$.

**Results of phase shift analysis**

The above presented methods make it possible to implement the phase shift analysis of the known experimental data for the elastic scattering differential cross sections of $n^3$He-channel in the energy interval 1-5 MeV [23-25]. Results of our first option are presented in Figs. 1-3. Value of mean square deviation $\chi^2$ was calculated according [23-25]. As for the computing details we refer to [15, 21]. Fig. 4 reported the differential cross sections from [23-25], and calculated curves obtained on the performed phase shift analysis done here.

It is well seen in Figs. 1-3 that *S* scattering phase shift has no resonances what is in accordance of the $^4$He observable spectra [26]. Resonance corresponding to the quantum numbers $J^\pi$, $T = 0^-,0$ at 21.01 MeV in *cm* is lying on the 0.575 MeV [26] higher the threshold of $n^3$He channel, i.e. 20.578 MeV associated with $^3P_0$ phase shift situated lower [23-25]. Resonances $2^-$, 0 at 21.84 MeV and $2^-$, 1 at 23.33MeV in *cm* are revealed at 1.68 MeV and 3.67 MeV in *ls,* but not in $^3P_2$ phase shift. Finally, two extra high resonances $1^-$, 1, and $1^-$, 0 at 23.64 MeV and 24.25 MeV at *cm* coordinating to $^3P_1$ or $^1P_1$ phase shifts do not reveal at 4.08 MeV and 4.90 MeV *ls* energies. We interpret this situation as the apart lying levels and not catching in present phase shift analysis.

In present analysis *S* phase shift starts from zero, but if potential contains both forbidden states (FS) and allowed states (AS), then it obeys to the generic Levinson theorem, and the corresponding phase shifts at zero energy should start from $n\pi\cdot$where *n* is a number of such states. In treating case there are two bound states, i.e. the ground one (GS) and first excited state (ES) lying at 20.21 MeV relatively to the GS. Thus, $^1S_0$ phase shift should start from 360 degrees. This variant we will refer as the second one, the corresponding results are given in Figs. 5 and 6, and Fig. 7 shows the conforming differential cross sections. Concluding, we should remark that considered case of *S* and *P* scattering waves is quite enough to reproduce the available differential cross sections rather well at the treated energies. The estimated error for phase shifts is on 10% level (see Fig. 5), what corresponds to the error bars of the differential cross sections used in the analysis. For comparison dots in Fig. 5 note the singlet $p^3$H scattering phase shifts from [27].

**Conclusion**

Therefore, the set of phase shifts for $n^3$He elastic scattering has been obtained basing on the analysis of experimental differential cross sections in the energy range from 1.0 up to 5.0 MeV. Within this set we succeeded in the reproducing of the differential angular distributions both quantitatively and qualitatively, what might be interesting while solving some problems of nuclear astrophysics [3]. Let stress once more, the obtained phase shifts are the input data for the constructing of the corresponding interaction potentials. These potentials in its turn are the standpoint for the calculations of reactions important for the astrophysical applications, for example, partially treated in our works [3,8].

Just omitting the details of the method of constructing such a potentials let us remark only that $p^3$H and $n^3$He systems are mixed by isospin, as they have the projection $T_z = 0$, and hence the total isospin values $T = 0$ and 1 are allowed. In these systems triplet and singlet phase shifts depend on two isospin values, and thus potentials depend effectively also [3]. Isospin mixing leads consequently to the mixing of the orbital states by Young schemes. In particular, it is kwon that in singlet spin state two orbital Young schemes allowed, i. e. {31} and {4} [4].

In [2] it was shown that mixed by isospin singlet $p^3$H scattering phase shifts may be presented as a half-sum of pure by isospin singlet phase shifts

$$\delta^{\{T=1\}+\{T=0\}} = 1/2\delta^{\{T=1\}} + 1/2\delta^{\{T=0\}},$$

that is equivalent to the analogous expression where the Young schemes are pointed

$$\delta^{\{4\}+\{31\}} = 1/2\delta^{\{31\}} + 1/2\delta^{\{4\}}.$$

Pure phase shifts with Young scheme {31} correspond to $T = 1$, and phase shifts with {4} to $T = 0$. As far as $p^3$He with $T_z = 1$ is pure by isospin $T = 1$, then it follows from given above expressions that basing on these kwon pure phase shifts (see, for example, [28]) and mixed $p^3$H phase shifts with possible $T = 0$ and 1 (see, for example, [29]) pure by isospin $T = 0$ phase shifts for $p^3$H channel can be constructed, as well as corresponding interaction potentials [2]. This approach is implying that pure phase shifts with $T = 1$ in $p^3$H channel make them comparable to the phase shifts with $T = 1$ in $p^3$He channel. Same procedure is valid for $n^3$He system while constructing both pure phase shifts and interaction potentials [2,3].

**References**

1. Dubovichenko S.B., Dzhazairov-Kakhramanov A.V. // Phys. Atom. Nucl. 1994. V. 57. №5. P. 733; Dubovichenko S.B., Dzhazairov-Kakhramanov A.V. // Phys. Part. Nucl. 1997. V. 28. P. 615.
2. Dubovichenko S.B., Neudatchin V.G., Sakharuk A.A. et al. // Izv. Akad. Nauk SSSR Ser. Fiz. 1990. V. 54. №5. P. 911; Neudatchin V.G., Sakharuk A.A., Dubovitchenko S.B. // Few Body Sys. 1995. V. 18. №2-4. P. 159; Dubovichenko S.B., Dzhazairov-Kakhramanov A.V. // Sov. J. Nucl. Phys. USSR 1990. V. 5 1. №6. P. 971; Dubovichenko S.B. // Phys. Atom. Nucl. 1995. V. 58. № 7. P. 1174.
3. Dubovichenko S.B. Thermonuclear processes of the Universe. New-York, NOVA Sci. Publ., 2012. 194p.; Dubovichenko S.B. Light nuclei and nuclear astrophysics. Sec. Edit., revised and expanded. Germany. Lambert Academy Publ. 2013. 320p.; Dubovichenko S.B. Selected method of nuclear astrophysics. Third Edit., revised and expanded. Germany. Lambert Academy Publ. 2013. 480 p.
4. Nemets O.F., Neudatchin V.G., Rudchik A.T., Smirnov Y.F., Tchuvil'sky Yu.M. Nucleon association in atomic nuclei and the nuclear reactions of the many nucleons transfers. Kiev: Naukova dumka. 1988. 488p. (in Russian); Neudatchin V.G., Sakharuk A.A., Smirnov Yu.F. Generalized potential description of interaction of the lightest cluster scattering and photonuclear reactions // Sov. J. Part. Nucl. 1992. V.23. P.210-271; Dubovichenko S.B., Uzikov Yu.N. // Phys. Part. Nucl. 2011. V. 42. №2. P. 251.

5. Neudatchin V.G. et al. // Phys. Rev. 1992. V .C45. P. 1512; Dubovichenko S.B., Zhusupov M.A. // Izv. Akad. Nauk SSSR Ser. Fiz. 1984. V .48. №5. P. 935; Dubovichenko S.B., Zhusupov M.A. // Sov. J. Nucl. Phys. USSR 1984. V. 39. №6. P. 870.

6. Dubovichenko S.B. // Phys. Atom. Nucl. 1998. V .61. №2. P. 162; Kukulin V.I., Pomerantsev V.N., Cooper S.G., Dubovichenko S.B. // Phys. Rev. 1998. V. C57. №5. P. 2462.

7. Dubovichenko S.B., Dzhazairov-Kakhramanov A.V. // Eur. Phys. Jour. 2009. V. A39. №2. P. 139; Dubovichenko S.B. // Phys. Atom. Nucl. 2010. V. 73. №9. P.1526; Dubovichenko S.B. // Russ. Phys. J. 2012. V.55. №2. P.138; Dubovichenko S.B. // Russ. Phys. J. 2011. V. 54. №2. P. 157; Dubovichenko S.B., Dzhazairov-Kakhramanov A.V. // Russ. Phys. J. 2009. V. 52. №8. P. 833; Dubovichenko S.B., Dzhazairov-Kakhramanov A.V. // Annal. der Phys. 2012. V. 524. №12. P. 850.

8. Dubovichenko S.B. // Phys. Part. Nucl. 2013. V. 44. №5. P. 803; Dubovichenko S.B., Dzhazairov-Kakhramanov A.V., Afanasyeva N.V. // Int. J. Mod. Phys. 2013. V. E22. №10. P. 1350075; Dubovichenko S.B., Dzhazairov-Kakhramanov A.V., Burkova N.A. // Int. J. Mod. Phys. 2013. V. E22. №5. P. 1350028; Dubovichenko S.B., Dzhazairov-Kakhramanov A.V. // Int. J. Mod. Phys. 2012. V. E21. №3. P. 1250039.

9. Neudatchin V.G., Smirnov Yu.F. Nucleon associations in light nuclei. Moscow: Nauka. 1969. 414p. (in Russian).

10. Angulo C. et al. // Nucl. Phys. 1999. V. A656. P. 3.

11. Adelberger E.G. et al. // Rev. Mod. Phys. 2011. V. 83. P.195.

12. Dubovichenko S.B., Zazulin D.M. // Russ. Phys. J. 2010. V. 53. №5. P. 458.

13. Dubovichenko S.B. // Russ. Phys. J. 2012. V. 55 №5. P. 561.

14. Dubovichenko S.B. // Russ. Phys. J. 2008. V. 51. №11. P. 1136.

15. Dubovichenko S.B. // Phys. Atom. Nucl. 2008. V.7 1. №1. P. 65.

16. Dubovichenko S.B. // Russ. Phys. J. 2009. V. 52. №7. P. 715.

17. Dubovichenko S.B. // Phys. Atom. Nucl. 2012. V. 75. №3. P.2 85.

18. Dubovichenko S.B. // Russ. Phys. J. 2014. (in print)

19. Dubovichenko S.B. // Russ. Phys. J. 2013. V. 55. №9. P. 992.

20. Tombrello T.A., Jones C.M., Phillips G.C., Weil J.L. // Nucl. Phys. 1962. V. 39. P. 541.

21. Dubovichenko S.B. Calculation methods of nuclear characteristics. Sec. Edit., revised and expanded. Germany. Lambert Academy Publ. 2012. 425 p.

22. Dubovichenko S.B. // Bull. Kaz.NTU (Almaty) 2004. №3. P.137.

23. Seagrave J.D., Cranberg L., Simmons J.E. // Phys. Rev. 1960. V. 119. P.19 81.

24. Sayres A.R., Jones K.W., Wu C.S. // Phys. Rev. 1961. V. 122. P. 1853.

25. Haesner B. et al. // Phys. Rev. 1983. V. C28. P .995.

26. Tilley D.R., Weller H.R., Hale G.M. // Nucl. Phys. 1992. V. A541. P. 1.

27. Kankowsky R. et al. // Nucl. Phys. 1976. V. A263. P. 29-46.

28. Tombrello T.A. // Phys. Rev. 1965. V.138. P.B40.

29. McIntosh J.S., Gluckstern R.L., Sack S. // Phys. Rev. 1952. V. 88. P. 752.

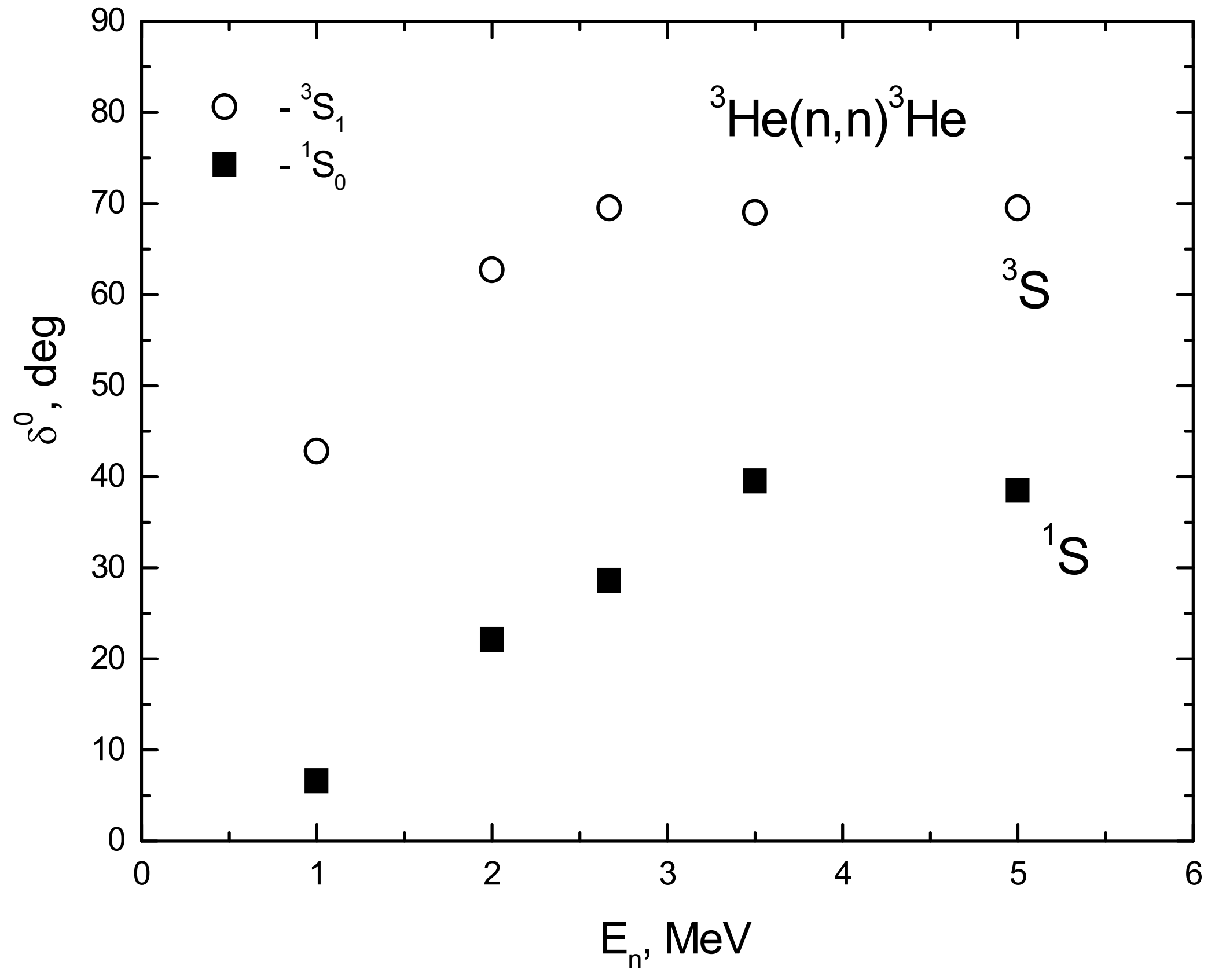

**Fig.1.** Elastic scattering $n^3$He phase shifts at low energies.

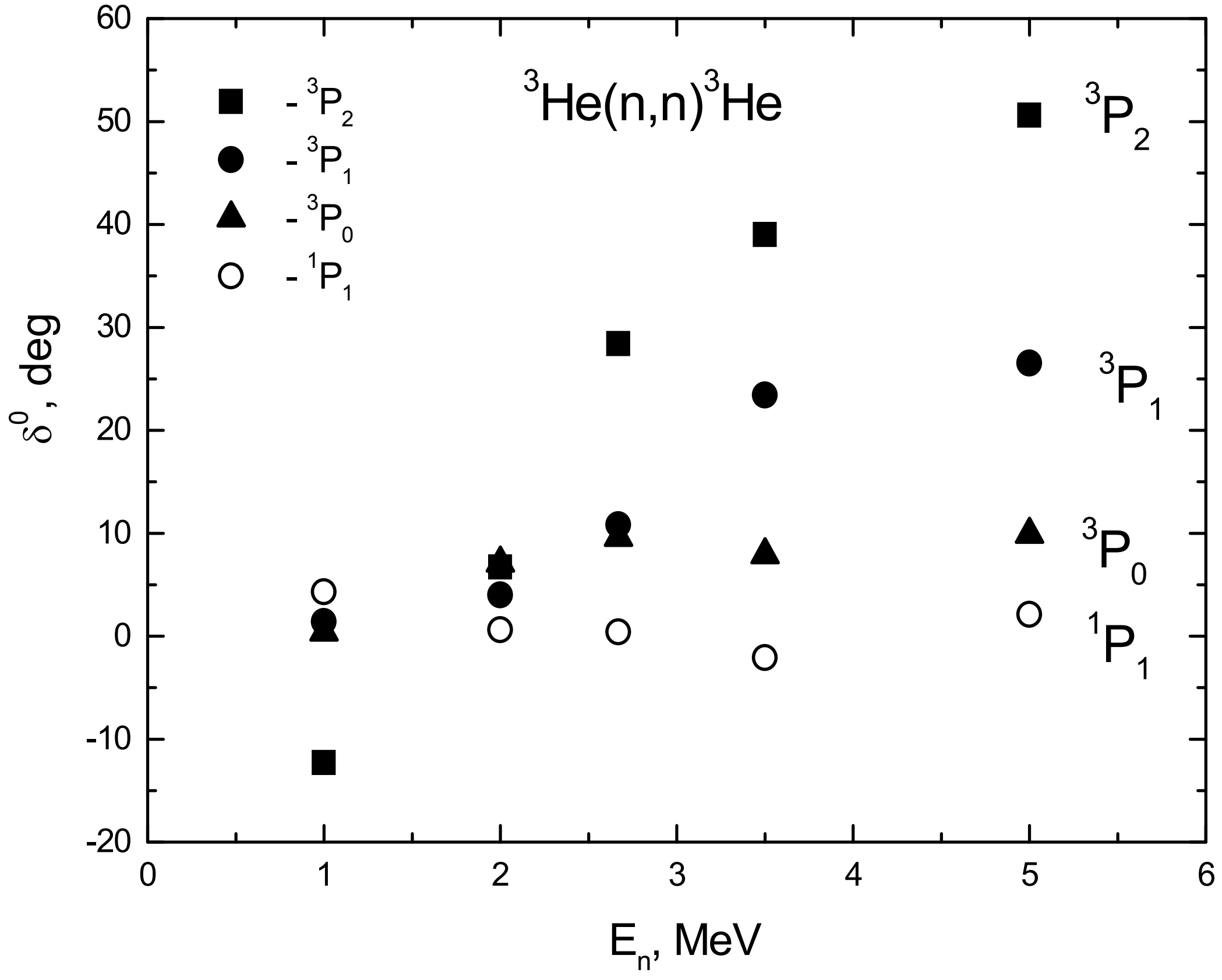

**Fig.2.** Elastic scattering $n^3$He phase shifts at low energies.

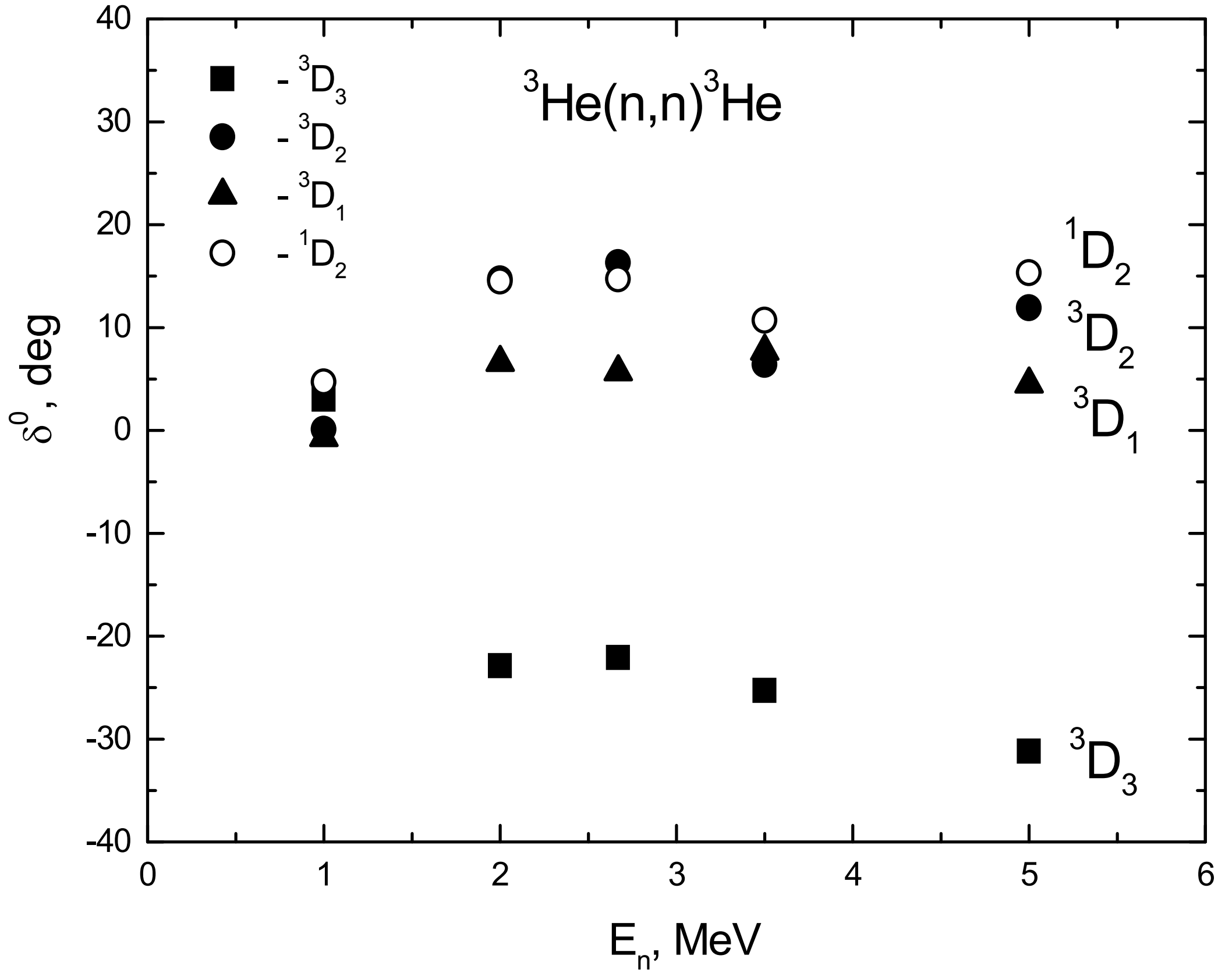

**Fig.3.** Elastic scattering $n^3$He phase shifts at low energies.

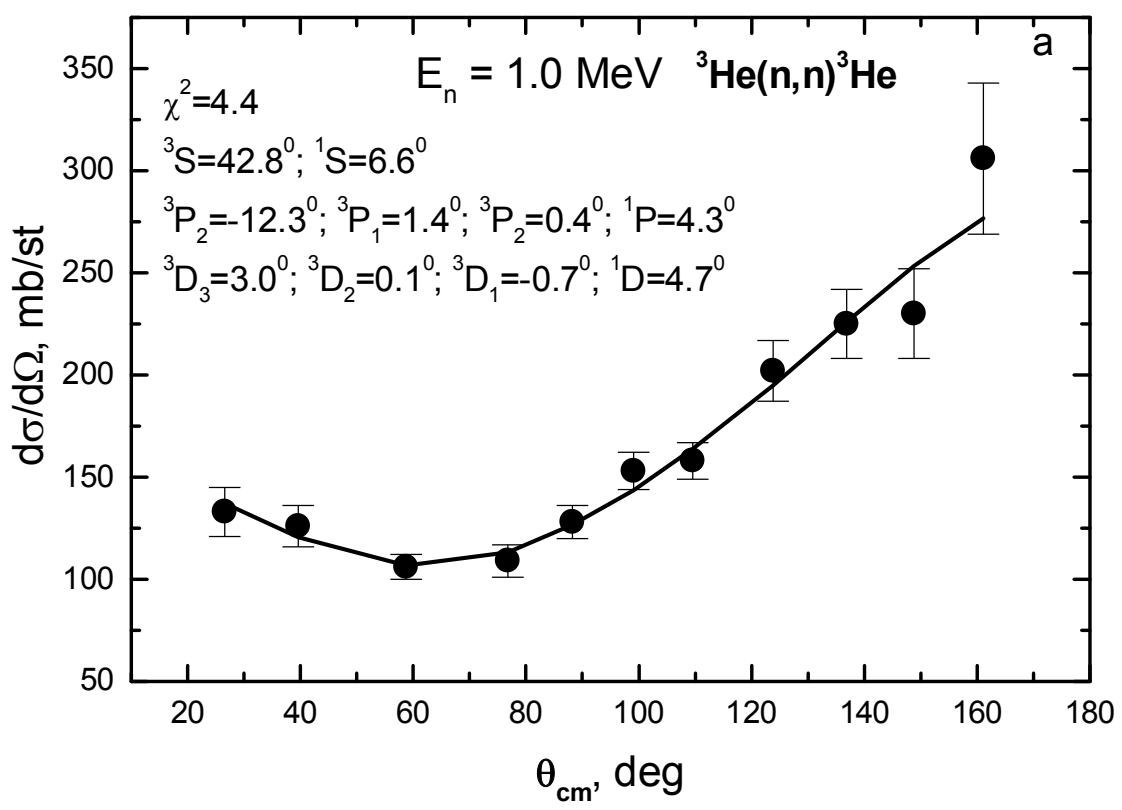

Experimental data [23].

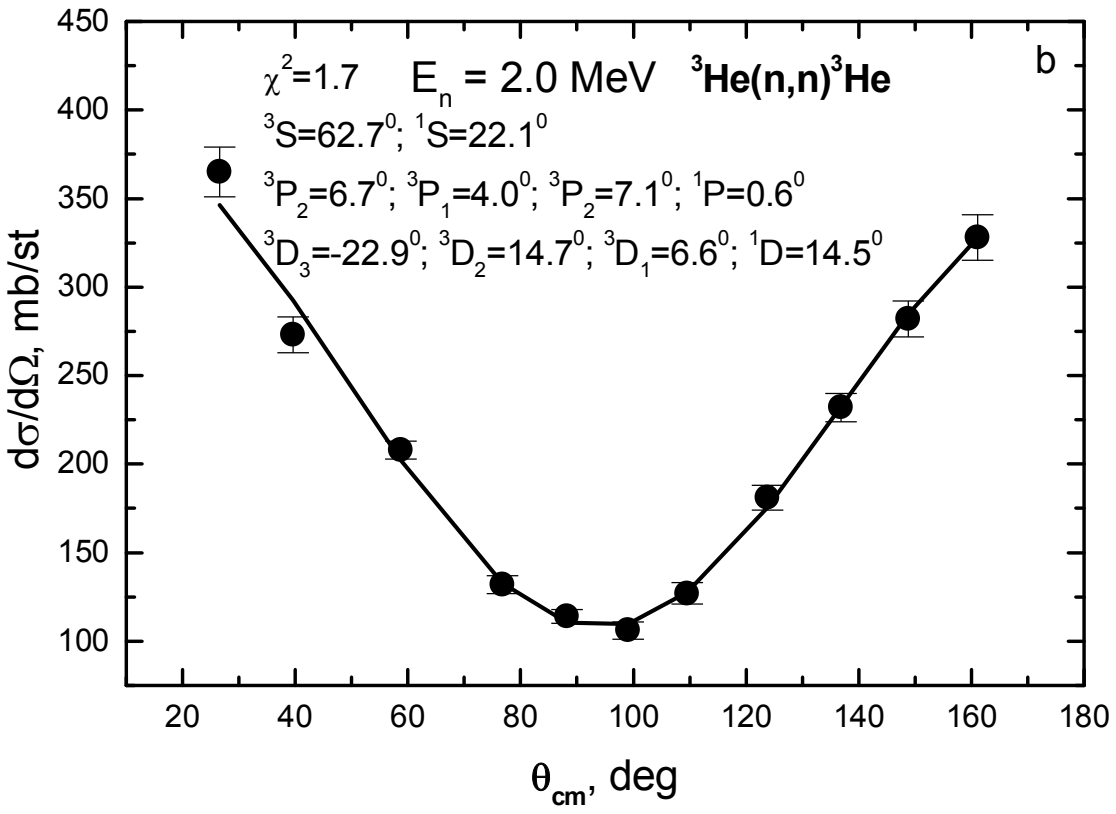

Experimental data [23].

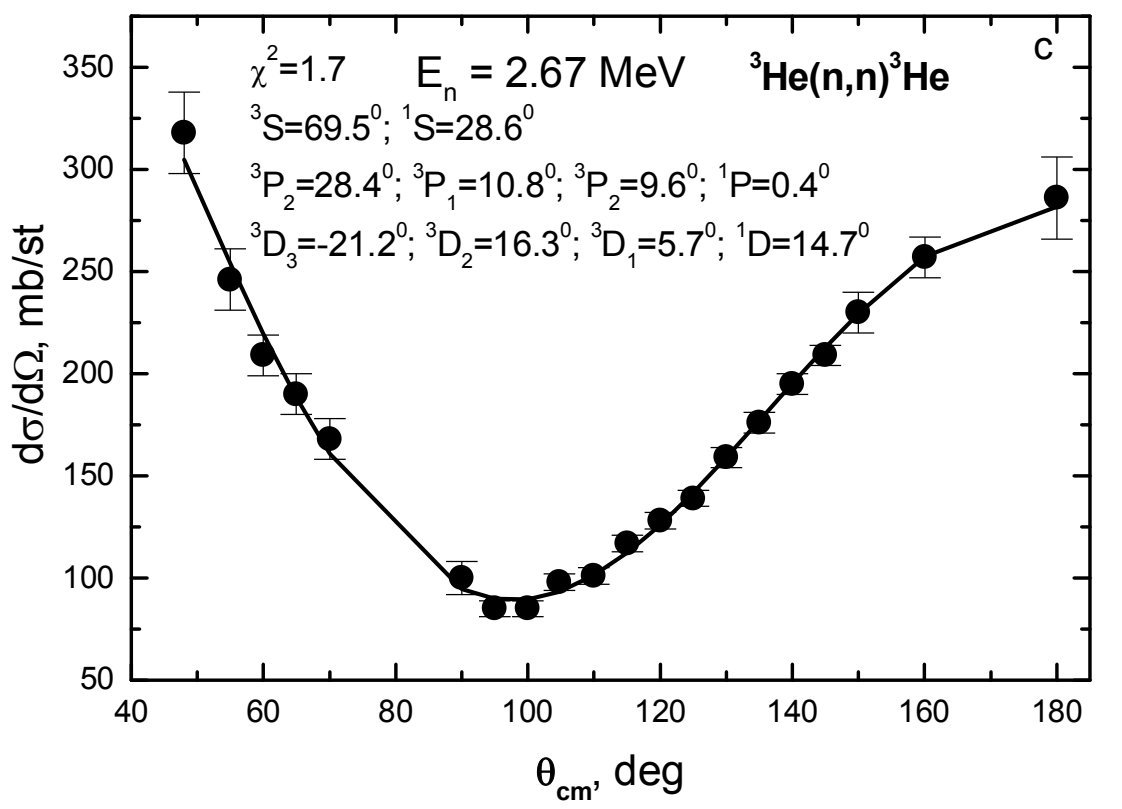

Experimental data [24].

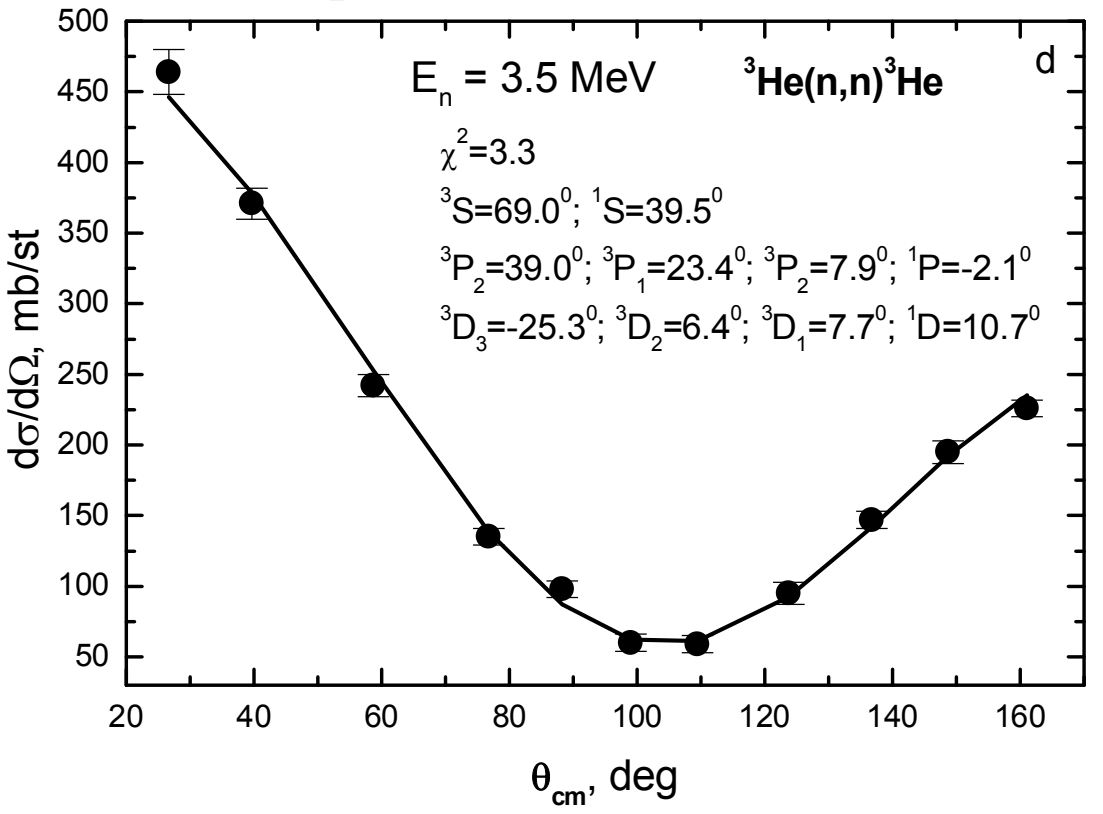

Experimental data [23].

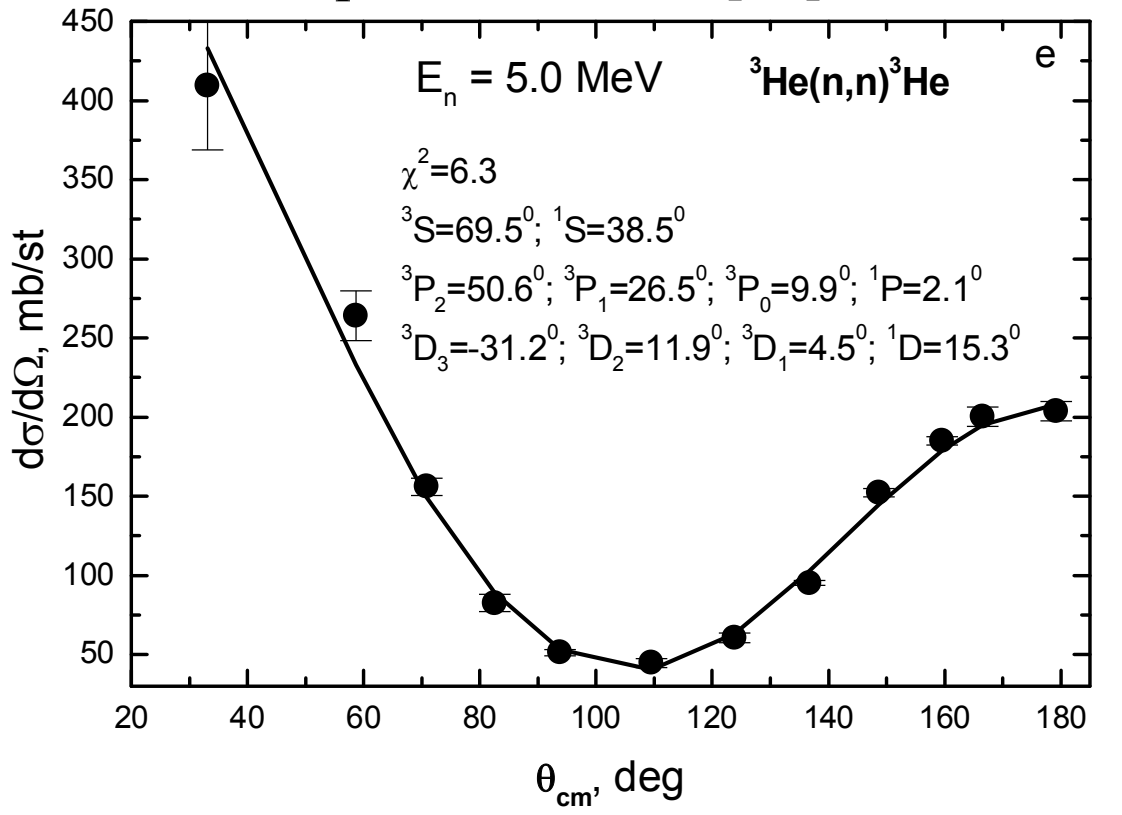

Experimental data [25].

**Fig.4.** Differential cross sections of $n^3$He elastic scattering.

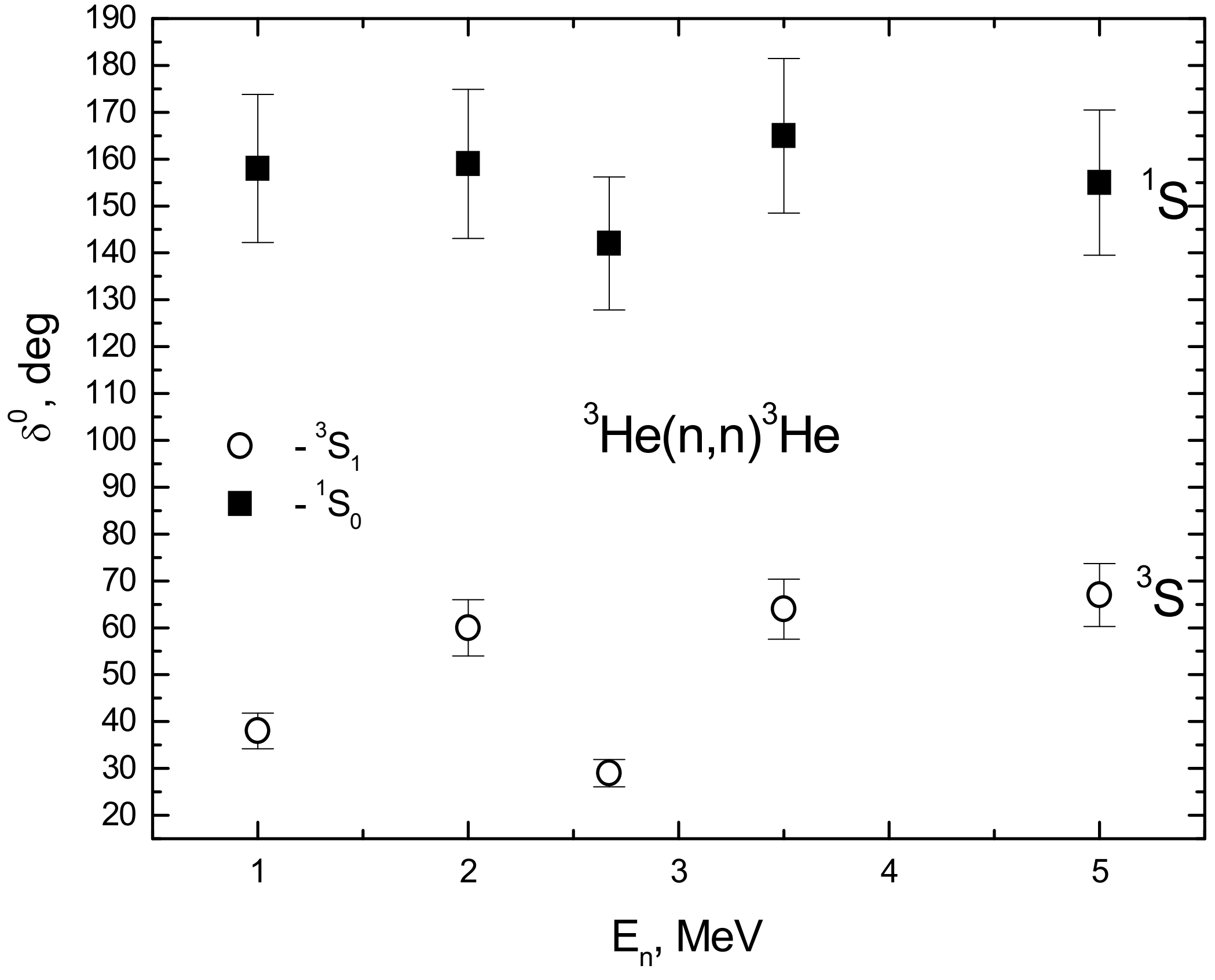

**Fig. 5.** Elastic scattering $n^3$He phase shifts at low energies.

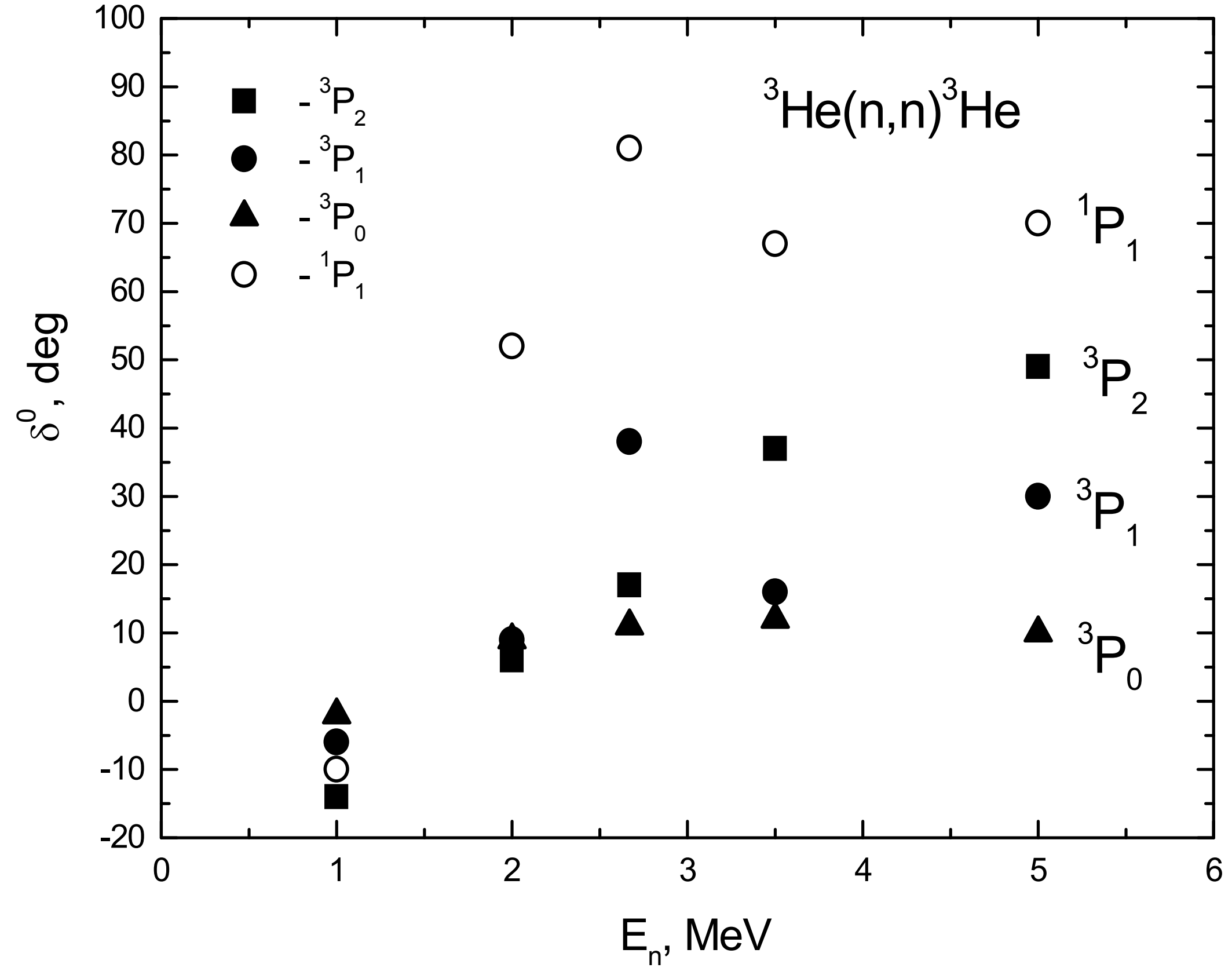

**Fig. 6.** Elastic scattering $n^3$He phase shifts at low energies.

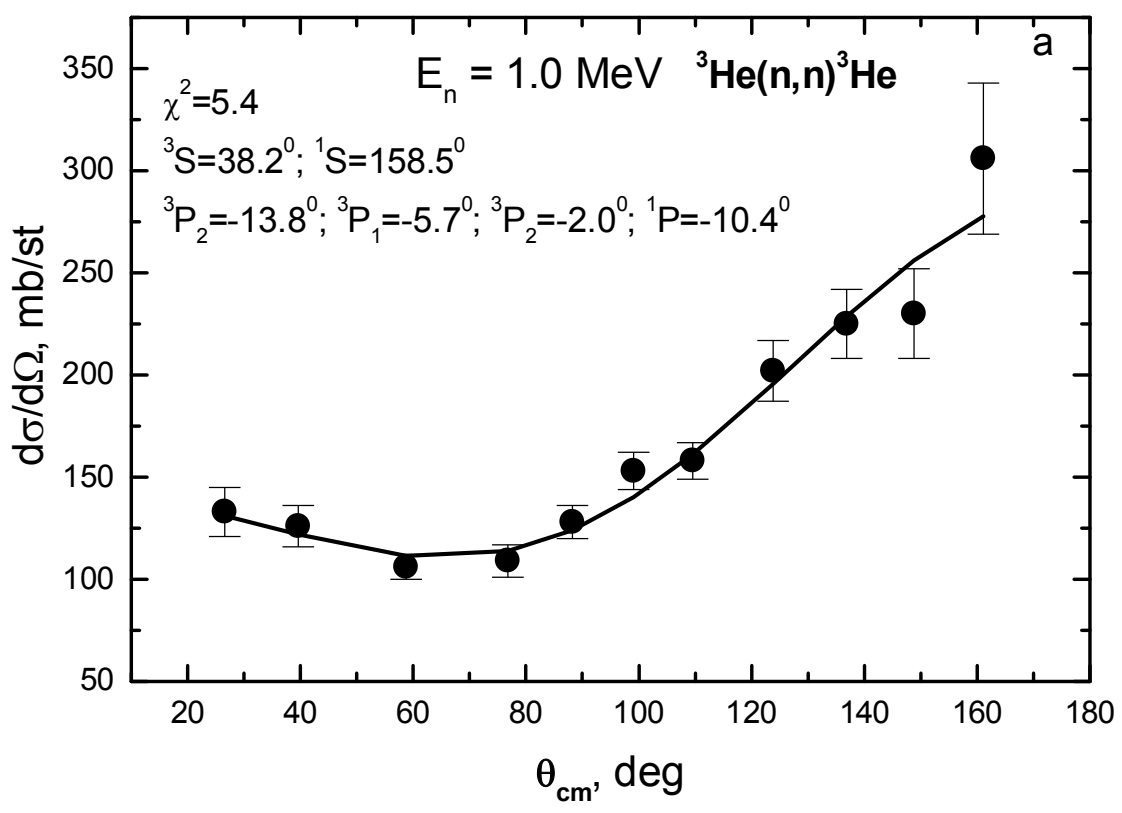

Experimental data [23].

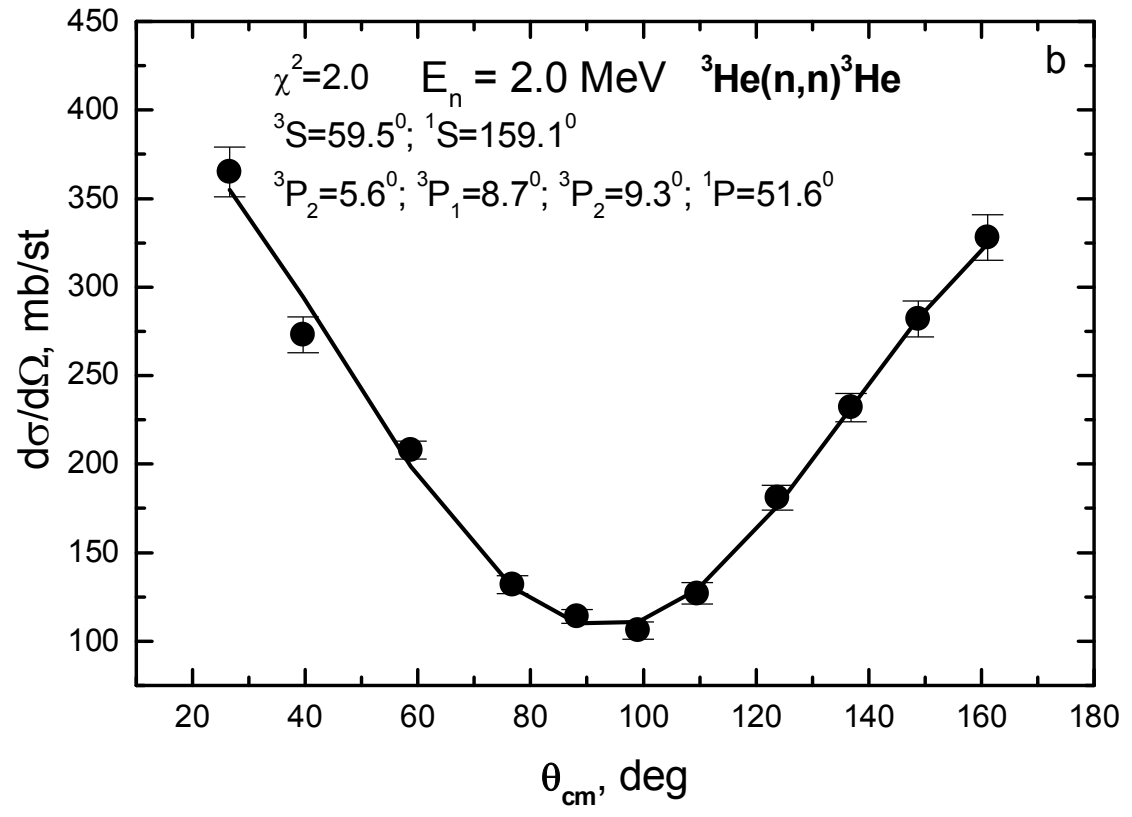

Experimental data [23].

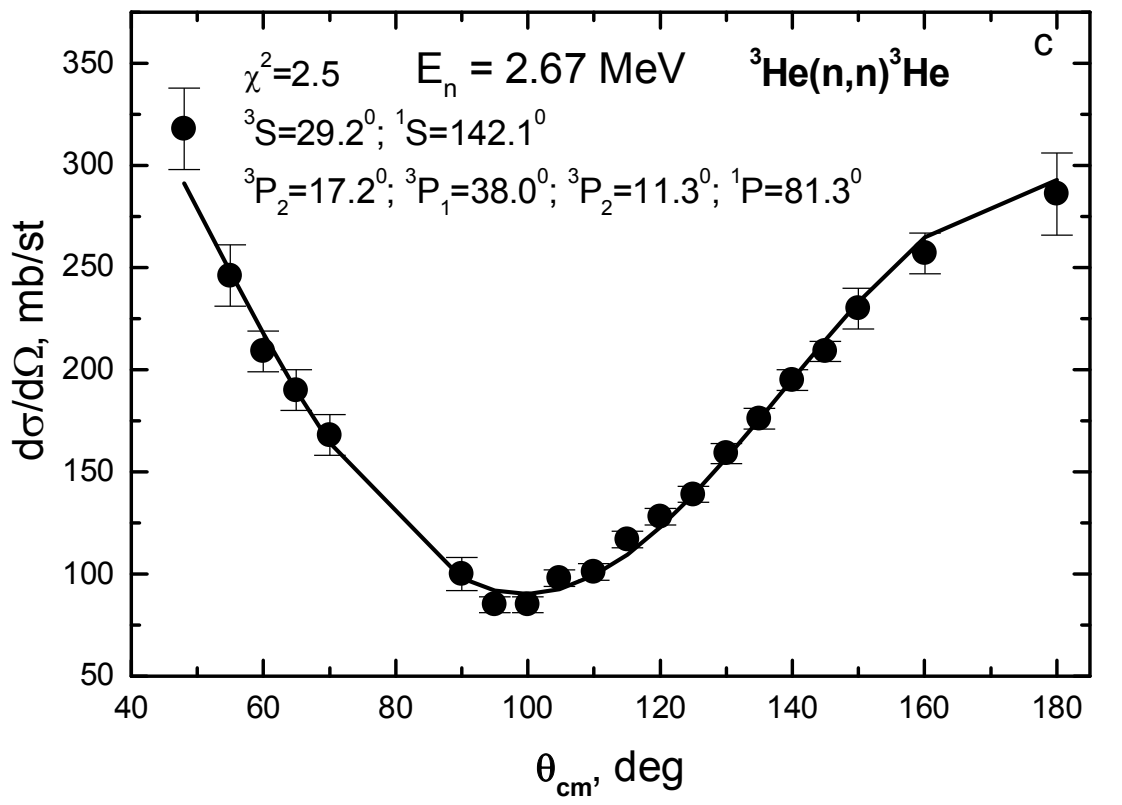

Experimental data [24].

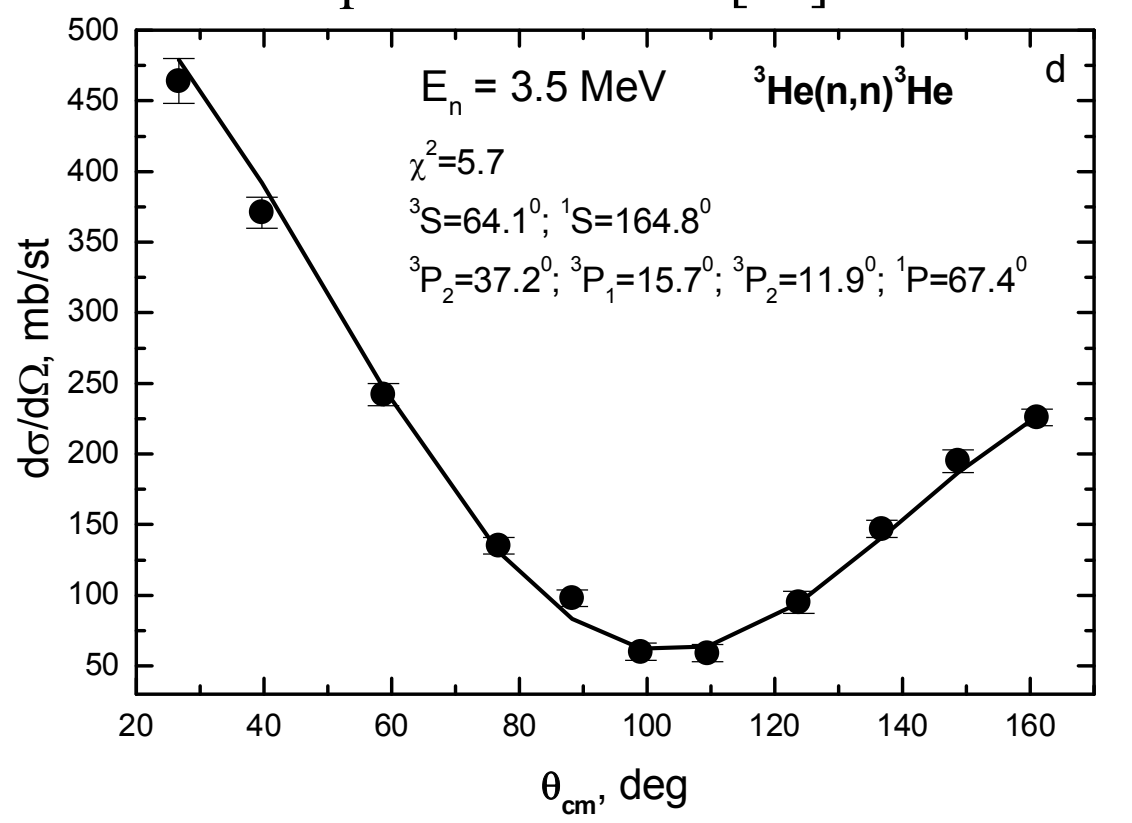

Experimental data [23].

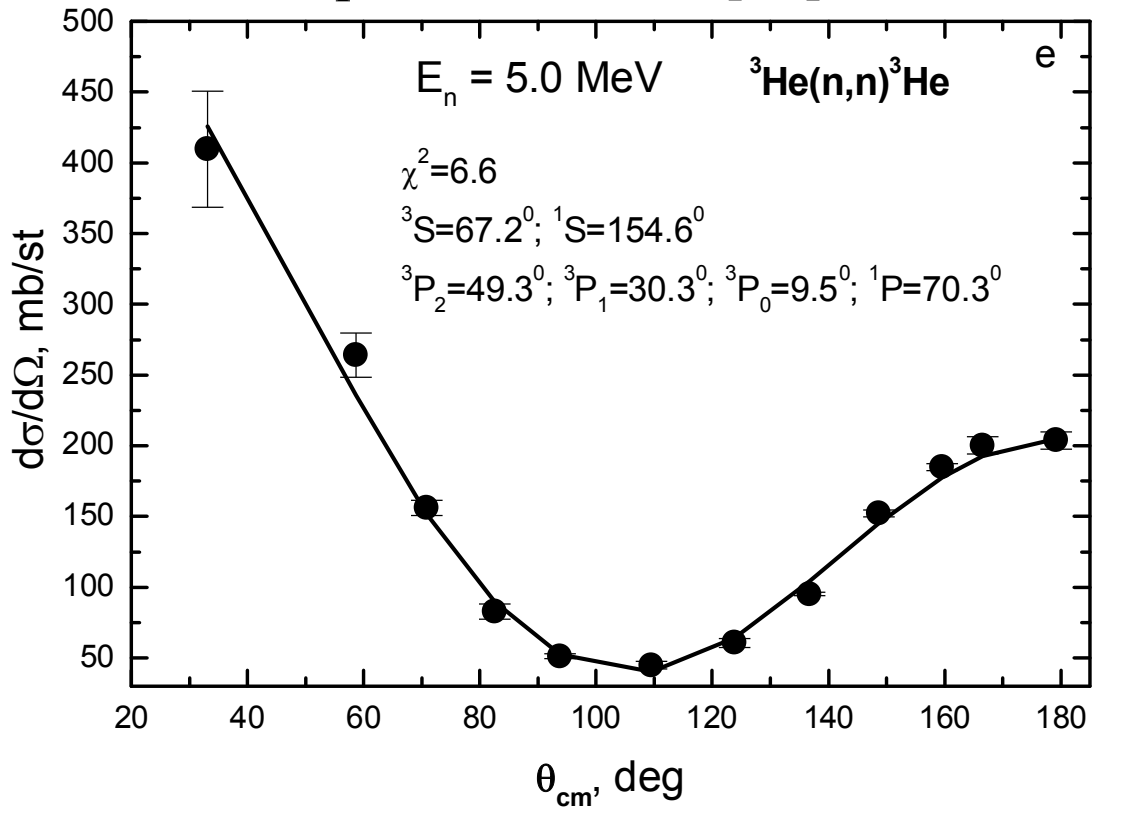

Experimental data [25].

**Fig. 7.** Differential cross sections of $n^3$He elastic scattering.